\documentclass[11pt,a4paper]{article}
\begin{titlepage}
\title{CDJ formulation from the instanton representation of Plebanski gravity}
\author{Eyo Eyo Ita III}
\date
\medskip  
\input amssym.def
\input amssym.tex

\def \in{\indent}

\begin{document}  
\maketitle
\bigskip
\centerline{Physics Department, US Naval Academy} 
\smallskip
\centerline{Annapolis, Maryland}
\smallskip
\centerline{ita@usna.edu}

\begin{abstract}
We show that a certain action which gives rise to the pure spin connection formulation of gravity by CDJ can be consistently derived from the action for the instanton representation of Plebanski gravity.  This is an illustrative example of when certain symmetries of the basic fields commute with the symmetries of the equations of motion.
\end{abstract}
\end{titlepage}

\section{CDJ formulation from the instanton representation}
Generally, when one reduces degrees of freedom of the basic variables at the level of an action $I$ one must be careful to ensure that one is not restricting the space of solutions, in order that the reduced action $I_0$ still correspond to the starting theory.  Inconsistencies can result when one makes gauge-fixing choices at the level of the action, or when one eliminates fields from an action using the equations of motion for other, different fields.  There is a certain action forming an intermediate step in the derivation of the pure spin connection formulation for gravity \cite{SPINCON} by Jacobson, Dell and Capovilla given by 
\begin{eqnarray}
\label{GEEVEN}
I_0=\int_M\bigl(\Psi_{ae}{F^a}\wedge{F^e}-\eta\bigl(\Lambda+\hbox{tr}\Psi^{-1}\bigr)\bigr),
\end{eqnarray}
\noindent
where $F^a={1 \over 2}F^a_{\mu\nu}{dx^{\mu}}\wedge{dx^{\nu}}$ is a $SO(3,C)$ curvature two form, $\Psi_{ae}\in{SO}(3,C)\times{SO}(3,C)$ a three by three matrix, and $\Lambda$ is the cosmological constant.  The action $I_0$ was obtained by elimination of certain auxiliary fields from Plebanski's action \cite{PLEBANSKI}.  In the present paper we will show that there is another action which consistently gives rise to $I_0$ through the elimination of auxiliary fields.  This action is referred to as the instanton representation of Plebanski gravity (IRPG), derived from Plebanski's action in \cite{EYO}.\par 
\indent 
The basic variables for the IRPG are a $SO(3,C)$ gauge connection $A^a_{\mu}$ and matrix $\Psi_{ae}\in{SO}(3,C)\otimes{SO}(3,C)$.\footnote{For index conventions, internal $SO(3,C)$ indices are labelled by symbols from the beginning of the Latin alphabet $a,b,c,\dots$, and spatial indices $i,j,k$ from the middle.  Greek symbols $\mu,\nu,\dots$ denote 4-dimensional spacetime indices.}
The action for the instanton representation of Plebanski gravity is given by
\begin{eqnarray}
\label{CANON9}
I_{Inst}=\int{dt}\int_{\Sigma}d^3x\biggl[\Psi_{ae}B^i_e\bigl(F^a_{0i}-\epsilon_{ijk}B^j_aN^k\bigr)\nonumber\\
-N(\hbox{det}B)^{1/2}\sqrt{\hbox{det}\Psi}\bigl(\Lambda+\hbox{tr}\Psi^{-1}\bigr)\biggr],
\end{eqnarray}
\noindent
where $B^i_a={1 \over 2}\epsilon^{ijk}F^a_{jk}$ and $F^a_{0i}$ are the spatial and temporal components of the $SO(3,C)$ field strength $F^a_{\mu\nu}$, and $N$ and $N^k$ are auxiliary fields (e.g. the lapse function and shift vector of metric General Relativity).  Also, we will assume $(\hbox{det}\Psi)$ and $(\hbox{det}B)$ are nonzero, which restricts the validity of (\ref{CANON9}) to Petrov Type I, D and O spacetimes.  The equation of motion for $\Psi_{ae}$ is given by
\begin{eqnarray}
\label{CANON10}
{{\delta{I}_{Inst}} \over {\delta\Psi_{ae}}}
=B^i_e\bigl(F^a_{0i}-\epsilon_{ijk}B^j_aN^k\bigr)
+N(\hbox{det}B)^{1/2}\sqrt{\hbox{det}\Psi}(\Psi^{-1}\Psi^{-1})^{ea}=0.
\end{eqnarray}
\noindent
Note, defining $\epsilon^{0ijk}\equiv\epsilon^{ijk}$ and using the relation 
\begin{eqnarray}
\label{CANON12}
B^i_{(e}F^{a)}_{0i}={1 \over 2}\epsilon^{ijk}F^{(e}_{jk}F^{a)}_{0i}
={1 \over 8}F^a_{\mu\nu}F^e_{\rho\sigma}\epsilon^{\mu\nu\rho\sigma},
\end{eqnarray}
\noindent
that the symmetric part of the equation of motion (\ref{CANON10}) is given by
\begin{eqnarray}
\label{CANON13}
{1 \over 8}F^a_{\mu\nu}F^e_{\rho\sigma}\epsilon^{\mu\nu\rho\sigma}+N(\hbox{det}B)^{1/2}\sqrt{\hbox{det}\Psi}(\Psi^{-1}\Psi^{-1})^{(ea)}=0.
\end{eqnarray}
For the antisymmetric part we simply contract (\ref{CANON10}) with $\epsilon_{dae}$ which gives
\begin{eqnarray}
\label{CANON14}
\epsilon_{dea}B^i_eF^a_{0i}=\epsilon_{ijk}\epsilon_{dea}B^i_eB^j_aN^k=2(\hbox{det}B)N^k(B^{-1})^d_k.
\end{eqnarray}
\noindent
This enables us to solve for the auxiliary field $N^i$ as
\begin{eqnarray}
\label{CANON15}
N^k={1 \over 2}\epsilon^{kij}F^a_{0i}(B^{-1})^a_j.
\end{eqnarray}
\noindent
It will be useful to write the Lagrangian (\ref{CANON9}) in covariant from by separating $\Psi_{ae}$ into its symmetric and antisymmetric parts $\Psi_{ae}=\Psi_{(ae)}+{1 \over 2}\epsilon_{aed}\psi_d$, where $\psi_d=\epsilon_{dae}\Psi_{ae}$ (hence $\Psi_{[ae]}={1 \over 2}\epsilon_{aed}\psi_d$).  Note that the integrand of the $N^k$ term in (\ref{CANON9}) can be written as $\epsilon_{ijk}N^iB^j_eB^k_a\Psi_{ae}=(\hbox{detB})N^i(B^{-1})^d_i\psi_d$.  Hence the action (\ref{CANON9}) can be written as
\begin{eqnarray}
\label{CANON17}
I_{Inst}=\int_Md^4x\biggl[{1 \over 8}\Psi_{ae}F^a_{\mu\nu}F^e_{\rho\sigma}\epsilon^{\mu\nu\rho\sigma}\nonumber\\
+\bigl({1 \over 2}\epsilon_{dae}F^a_{0i}B^i_e+N^i(B^{-1})^d_i\bigr)\psi_d
-N(\hbox{det}B)^{1/2}\sqrt{\hbox{det}\Psi}\bigl(\Lambda+\hbox{tr}\Psi^{-1}\bigr)\biggr].
\end{eqnarray}
\noindent
The equation of motion for $N^i$ implies that $\psi_d=0$.  But since $\psi_d$ is also an independent dynamical field, then it is permissible to set $\psi_d=0$ only after, not before, writing down its Lagrange equation of motion 
\begin{eqnarray}
\label{CANON18}
{{\delta{I}_{Inst}} \over {\delta\psi_d}}\biggl\vert_{\psi_d=0}={1 \over 2}\epsilon_{dae}F^a_{0i}B^i_e+N^i(B^{-1})^d_i(\hbox{det}B)=0.
\end{eqnarray}
\noindent
Similarly, the equation of motion for $N$ is equivalent to $\Lambda+\hbox{tr}\Psi^{-1}=0$.  The solution to (\ref{CANON18}) is given precisely by (\ref{CANON15}).  The result is that the antisymmetric part of the equation of motion for $\Psi_{ae}$ is the same as the equation of motion for the antisymmetric part of $\Psi_{ae}$.\par 
\indent
To find the equation of motion for the connection $A^a_{\mu}$ it will be convenient to use the following relation $\epsilon_{ijk}N^iB^j_eB^k_a\Psi_{[ae]}={1 \over 2}\epsilon_{ijk}N^iB^j_aB^k_e\epsilon_{aed}\psi_d$.  Then the action (\ref{CANON17}) can also be written as
\begin{eqnarray}
\label{CANON19}
I_{Inst}=\int_Md^4x\biggl[{1 \over 8}\Psi_{ae}F^a_{\mu\nu}F^e_{\rho\sigma}\epsilon^{\mu\nu\rho\sigma}
+{1 \over 2}\epsilon_{dae}\bigl(F^a_{0j}B^j_e-\epsilon_{ijk}N^iB^j_eB^k_a\bigr)\psi_d\nonumber\\
-N(\hbox{det}B)^{1/2}\sqrt{\hbox{det}\Psi}\bigl(\Lambda+\hbox{tr}\Psi^{-1}\bigr)\biggr].
\end{eqnarray}
\noindent
The equation of motion for $A^a_{\mu}$ can be found by integration by parts of all terms containing the connection, which yields
\begin{eqnarray}
\label{CANON20}
{{\delta{I}} \over {\delta{A}^a_{\mu}}}=-\epsilon^{\mu\nu\rho\sigma}D_{\nu}(\Psi_{ae}F^e_{\rho\sigma})
-{1 \over 2}\delta^{\mu}_i\epsilon^{jml}D_m\biggl[\epsilon_{dag}\bigl(F^a_{0j}-2\epsilon_{jki}B^k_aN^i\bigr)\psi_d\nonumber\\
+(B^{-1})^g_jN(\hbox{det}B)^{1/2}\sqrt{\hbox{det}\Psi}\bigl(\Lambda+\hbox{tr}\Psi^{-1}\bigr)\biggr]=0.
\end{eqnarray}
\noindent
Note, on solution to the $N$ and $N^k$ equations, that the terms in large brackets in (\ref{CANON20}) vanish.  Therefore on-shell, (\ref{CANON20}) reduces to 
\begin{eqnarray}
\label{CANON21}
{{\delta{I}} \over {\delta{A}^a_{\mu}}}=\epsilon^{\mu\rho\sigma\nu}F^e_{\rho\sigma}D_{\nu}\Psi_{(ae)}=0,
\end{eqnarray}
\noindent
where we have used the Bianchi identity $\epsilon^{\mu\nu\rho\sigma}D_{\nu}F^e_{\rho\sigma}=0$.  Note that we can eliminate $\psi_d$ and $N^i$ by evaluating the action (\ref{CANON17}) on the critical point $\psi_d=0,N^i={1 \over 2}\epsilon^{ijk}F^a_{0j}(B^{-1})^a_k$, which leads to 
\begin{eqnarray}
\label{CANON22}
I_0=I_{Inst}\biggl\vert_{\psi_d=0,N^i={1 \over 2}\epsilon^{ijk}F^a_{0j}(B^{-1})^a_k}\nonumber\\
=\int_Md^4x\biggl[{1 \over 8}\Psi_{ae}F^a_{\mu\nu}F^e_{\rho\sigma}\epsilon^{\mu\nu\rho\sigma}+\eta\bigl(\Lambda+\hbox{tr}\Psi^{-1}\bigr)\biggr]
\end{eqnarray}
\noindent
where $\eta=\sqrt{\hbox{det}B}\sqrt{\hbox{det}\Psi}$, whence $\Psi_{ae}=\Psi_{(ae)}$ is now symmetric.  Note that the equation of motion of (\ref{CANON22}) is precisely the symmetric part of the equation of motion for $\Psi_{ae}$ in (\ref{CANON13}), and moreover that (\ref{CANON21}) can be seen as the equation of motion for $A^a_{\mu}$ from (\ref{CANON22}).  Additionally, the action (\ref{CANON22}) is precisely the same action as (\ref{GEEVEN}) which means the following things: (i) $I_0\subset{I}_{Inst}$, namely that $I_{Inst}$ is a different action from the one leading to the CDJ formalism, and contains this action as a subset (ii) the symmetries of the equations of motion commute with the symmetries of the Lagrangian $I_{Inst}$ in this case. (iii) In a certain interpretation, $N^i$ and $\psi_d$ can be unified into one six-component field $\Phi_{\alpha}$.  Then elimination of $\Phi_{\alpha}$ from $I_{Inst}$ through its equations of motion to get $I_0$ is a self-consistent procedure.


\begin{thebibliography}{99}

\bibitem{SPINCON} {R. Capovilla, J. Dell and T. Jacobson `A pure spin connection formulation of gravity'
Class. Quantum Grav. 8 (1991) 59-73}

\bibitem{PLEBANSKI} {Jerzy Plebanski `On the separation of Einsteinian substructures'
J. Math. Phys. Vol. 18, No. 2 (1977)}

\bibitem{EYO} {Eyo Ita `Instanton Representation of Plebanski Gravity. Gravitational Instantons from the Classical Formalism', The Abraham Zelmanov Journal, 2011, volume 4, pages 36-71}

\end{thebibliography}
\end{document}